\documentclass[runningheads]{llncs}
\usepackage[T1]{fontenc}
\usepackage{graphicx}

\usepackage[svgnames,table]{xcolor} %
\usepackage{setspace}%

\usepackage[ruled,vlined]{algorithm2e}
\usepackage{eurosym}
\usepackage{tikz}
\usepackage{pgfplots}

\usepackage{algpseudocode}

\usepackage[hidelinks]{hyperref}
\hypersetup{
    colorlinks,
    citecolor=blue,
    filecolor=blue,
    linkcolor=blue,
    urlcolor=blue
    }
\usepackage{color}

\usepackage{bbm}
\usepackage{amsmath,amsfonts,amssymb}

\usepackage{amsthm} %

\usepackage{xspace}
\usepackage{array}
\usepackage{url}
\usepackage{mathtools}
\usepackage{algorithm2e}
\usepackage{code}

\usepackage{tikz}
\usetikzlibrary{arrows,shapes,positioning}
\usetikzlibrary{decorations.markings}

\usepackage{makecell}

\usepackage{mathtools} %

\usepackage[utf8]{inputenc} %
\usepackage[svgnames,table]{xcolor} %
\usepackage{setspace}%
\usepackage[ruled,vlined]{algorithm2e}

\usepackage{csquotes}

\usepackage{graphicx} %
\usepackage{bbm}
\usepackage{amsmath,amsfonts,amssymb}
\usepackage{xspace}
\usepackage{array}
\usepackage{url}
\usepackage{mathtools}
\usepackage{algorithm2e}
\usepackage{code}

\usepackage{tikz}
\usetikzlibrary{arrows,shapes,positioning}
\usetikzlibrary{decorations.markings}
\usepackage{makecell}
\usepackage{mathtools} %

\usepackage{xspace}
\usepackage{graphicx}
\usepackage{amsfonts}
\usepackage{amsmath}
\usepackage{booktabs}

\usepackage{amsthm}

\newcommand{\connectedcomponent}{\mathcal{CC}\xspace}

\newcommand{\VECTEURAB}{\overrightarrow{ab}\xspace}

\newcommand{\opp}{\mathrm{Opp}\xspace}

\newcommand{\manifold}{\mathbb{M}\xspace}
\newcommand{\manifoldtwo}{\mathbb{N}\xspace}

\newcommand{\myfunction}{F\xspace}

\newcommand{\Zeals}{\mathbb{Z}\xspace}

\newcommand{\minima}{M_{-}\xspace}

\newcommand{\EXT}{\mathcal{G}^+\xspace}

\newcommand{\indicatortwo}{1_{\manifoldtwo}\xspace}
\newcommand{\indicatorST}{1_{\{\sigma,\tau\}}\xspace}
\newcommand{\myfunctiontwo}{\myfunction'\xspace}
\newcommand{\complex}{\mathbb{K}\xspace}
\newcommand{\gradient}{\overrightarrow{\mathrm{grad}}\xspace}

\newcommand{\graph}{\mathcal{G}\xspace}
\newcommand{\myfunctiongraph}{\myfunction_{\graph}\xspace}
\newcommand{\graphmanifold}{\graph_{\myfunction}\xspace}
\newcommand{\Edges}{\mathrm{Edges}\xspace}

\newcommand{\myproof}{\textbf{\underline{Proof:}}\xspace}

\newcommand{\DIM}{\mathrm{dim}\xspace}
\newcommand{\CLOSE}{\mathrm{Clo}\xspace}
\newcommand{\STAR}{\mathrm{St}\xspace}

\newtheorem{Th}[definition]{Theorem}
\newtheorem{Coro}[definition]{Corollary}
\newtheorem{Lem}[definition]{Lemma}
\newtheorem{Proposition}[definition]{Proposition}

\newtheorem{Cons[definition]}{Consequence}

\newcommand{\elimine}[1]{}

\pgfplotsset{compat=1.17} 

\providecommand{\dontprintsemicolon}{\DontPrintSemicolon}

\begin{document}
\title{Gradient Vector Fields of Discrete Morse Functions and Watershed-cuts}
\titlerunning{GVFs and watersheds}
\author{Nicolas Boutry\inst{1}\orcidID{0000-0001-6278-4638}
\and Gilles Bertrand\inst{2}
  \and\newline Laurent Najman\inst{2}\orcidID{0000-0002-6190-0235}
}
\authorrunning{N. Boutry, G. Bertrand, and L. Najman}
\institute{
  EPITA Research and Development Laboratory (LRDE), France
  \email{nicolas.boutry@lrde.epita.fr}\\
  Univ Gustave Eiffel, CNRS, LIGM, F-77454 Marne-la-Vallée, France
  \email{\{gilles.bertrand,laurent.najman\}@esiee.fr}
}%
\maketitle              %
\begin{abstract}
In this paper, we study a class of discrete Morse functions, coming from Discrete Morse Theory, that are equivalent to a class of simplicial stacks, coming from Mathematical Morphology. We show that, as in Discrete Morse Theory, we can see the gradient vector field of a simplicial stack (seen as a discrete Morse function) as the only relevant information we should consider. Last, but not the least, we also show that the Minimum Spanning Forest of the dual graph of a simplicial stack is induced by the gradient vector field of the initial function. This result allows computing a watershed-cut from a gradient vector field. 

\end{abstract}

\section{Introduction}

We present here several results relating Mathematical Morphology~\cite{najman2013mathematical} (MM) to Discrete Morse Theory~\cite{forman1995discrete} (DMT). This strengthens previous works highlighting links between MM and topology.  In~\cite{cousty2009collapses,cousty2014collapses}, it is demonstrated that watersheds are included in skeletons on pseudomanifolds of arbitrary dimension. Recently  (see \cite{boutry2019equivalence,boutry2021equivalence,boutry:hal-03676854}), some relations between MM and Topological Data Analysis~\cite{tierny2017introduction,munch2017user} (TDA)  have been exhibited: the \emph{dynamics}~\cite{grimaud1992new}, used in MM to compute markers for watershed-based image-segmentation, is equivalent to the \emph{persistence}, a fundamental tool from Persistent Homology~\cite{edelsbrunner2008persistent}.

In this paper, the first main result links the spaces used in MM and in TDA: the main mathematical spaces used in DMT,  \emph{discrete Morse functions}~\cite{scoville2019discrete} (DMF), are equivalent, under some constraints, to spaces well-known in MM and called \emph{simplicial stacks}~\cite{cousty2009watershed,cousty2009collapses,cousty2014collapses}. Simplicial stacks are a class of weighted simplicial complexes whose upper threshold sets are also complexes. Indeed, in a DMF, the values locally increase when we increase the dimension of the face we are observing; in a simplicial stack, it is the opposite. Without surprise, we can then observe that, under some constraints, any DMF is the opposite of a simplicial stack, and conversely.

In TDA, it is a common practice to consider that the main information conveyed by a DMF is its gradient vector field (GVF), naturally obtained by pairing neighbor faces with the same value. Two DMFs with the same GVF are then considered to be equivalent. Using the very same principle on simplicial stacks, we can go further, and consider that a GVF encodes not only a class of DMFs, but also the corresponding class of simplicial stacks.

The relation between TDA and MM in the context of DMFs and stacks is not limited to the previous observations. In \cite{cousty2009watershed}, the authors proved that a watershed-cut is a Minimum Spanning Forests (MSF) cut in the dual graph of a  simplicial stack. 
We prove here that such a MSF can be extracted from the GVF of the simplicial stack (seen as a DMF). Relations between watersheds and Morse theory have long been informally known \cite{de2015morse}, but this is the first time that a link is presented in the discrete setting, relying on a precise definition of the watershed. Furthermore, as far as we know, this is the first time that a concept from Discrete Morse Theory is linked to a classical combinatorial optimization problem.

The plan of this paper is the following. Section~\ref{sec.maths} recalls the mathematical background necessary to our proofs. Section~\ref{sec.equiv} shows the  equivalence between DMF's and simplicial stacks. Section~\ref{sec.theproof} studies the link between MSFs and GVFs.
Section~\ref{sec.conclusion} concludes the paper.

\section{Mathematical background}
\label{sec.maths}

\subsection{Simplicial complexes, graphs and pseudomanifolds}
We call \emph{(abstract) simplex} any finite nonempty set of arbitrary elements. The \emph{dimension} of a simplex $x$, denoted by $\DIM(x)$, is the number of its elements minus one. In the following, a simplex of dimension $d$ will also be called a \emph{$d$-simplex}. If $x$ is a simplex, we set $\CLOSE(x)=\{y| y\subseteq x, y \neq \emptyset\}$. A finite set $X$ of simplices is a \emph{cell} if there exists $x \in X$ such that $X = \CLOSE(x)$.

If $X$ is a finite set of simplices, we write $\CLOSE(X)= \cup\{\CLOSE(x) | x \in X\}$, the set $\CLOSE(X)$ is called the \emph{(simplicial) closure} of $X$. A finite set $X$ of simplices is a \emph{(simplicial) complex} if $X = \CLOSE(X)$. 

In the sequel of the paper, $\complex$ denotes a simplicial complex.  A subcomplex of $\complex$ is a subset of $\complex$ which is also a complex. Any element in $\complex$ is a face of $\complex$ and we call \emph{$d$-face} of $\complex$ any face of $\complex$ whose dimension is $d$. If $\sigma, \tau$ are two faces of $\complex$ with $\tau\subset\sigma$, we say that $\sigma$ is a coface of $\tau$. Any  $d$-face of $\complex$ that is not included in any $(d+1)$-face of $\complex$ is called a \emph{$(d-)$facet} of $\complex$ or a \emph{maximal face} of $\complex$. 

The dimension of $\complex$, written \emph{$\DIM(\complex)$}, is the largest dimension of its faces: $\DIM(\complex) = \max\{\DIM(x) | x \in \complex\}$, with the convention that $\DIM(\emptyset)=-1$. If $d$ is the dimension of $\complex$, we say that $\complex$ is \emph{pure} whenever the dimension of all its facets equals $d$.

Suppose that there is a pair of simplices\footnote{The superscripts correspond to the dimensions of the faces.} $(\sigma^{(p-1)},\tau^{(p)})$ of $\complex$ with $\sigma \subset \tau$ such that the only coface of $\sigma$ is $\tau$. Then $\complex \setminus \{\sigma,\tau\}$ is a simplicial complex called \emph{an elementary collapse} of $\complex$.  For an elementary collapse, such a pair $\{\sigma,\tau\}$ is called a \emph{free pair}, and $\sigma$ is called a {\em free face}. Note that elementary collapses preserve simple homotopy type \cite{whitehead1939simplicial}. A free pair $\{\sigma^{(d-1)},\tau^{(d)}\}$ is called a {\em free $d$-pair}, and $\complex \setminus \{\sigma^{(d-1)},\tau^{(d)}\}$ is called an {\em elementary $d$-collapse}. If a complex $\complex'$ is the result of a sequence of elementary $d$-collapses of $\complex$, we say that $\complex'$ is a {\em $d$-collapse of $\complex$}. If, furthermore, there is no free $d$-pair for $\complex'$, then $\complex'$ is an {\em ultimate $d$-collapse of $\complex$}.

In this paper, a {\em graph} $\graph$ is a pure 1-dimensional simplicial complex. A subgraph is a subset of a graph which is also a graph.  We denote the vertices (the 0-dimensional elements) of a graph $\graph$ by $V(\graph)$, and the edges (the 1-dimensional elements) by $E(\graph)$.

Let $X$ be a set of simplices, and let $d \in \mathbb{N}$. Let $\pi = \langle x_0, \ldots, x_l \rangle$ be a sequence of $d$-simplices in $X$. The sequence $\pi$ is a \emph{$d$-path} from $x_0$ to $x_l$ in $X$ if $x_{i-1} \cap x_i$ is a $(d-1)$-simplex in $X$, for any $i \in\{1, \ldots, l\}$. Two $d$-simplices $x$ and $y$ in $X$ are said to be \emph{$d$-linked} for $X$ if there exists a $d$-path from $x$ to $y$ in $X$. We say that the set $X$ is \emph{$d$-connected} if any two $d$-simplices in $X$ are $d$-linked for $X$. We say that the set $Y \subset X$ is a \emph{$d$-connected component} (or simply, a {\em connected component}) of $X$ if $Y$ is $d$-connected and maximal for this property.

Let $X$ be a set of simplices, and let $\pi = \langle x_0, \ldots ,x_l \rangle$ be a $d$-path in $X$. The $d$-path $\pi$ is said \emph{simple} if for any two distinct $i$ and $j$ in $\{0,\ldots,l\}$, $x_i \neq x_j$. It can be easily seen that $X$ is $d$-connected if and only if, for any two $d$-simplices $x$ and $y$ of $X$, there exists a simple $d$-path from $x$ to $y$ in $X$.

A complex $\complex$ of dimension $d$ is said to be a \emph{$d-$pseudomanifold} if 
\begin{itemize}
    \item[(1)] $\complex$ is pure,
    \item[(2)] any $(d - 1)-$face of $\complex$ is included in exactly two $d-$faces of $\complex$, and
    \item[(3)] $\complex$ is $d-$connected. 
\end{itemize}  
In the sequel of the paper, $d \geq 1$ is an integer, and $\manifold$ denotes a $d$-pseudo-manifold.

\begin{Proposition}[Ultimate collapses~\cite{cousty2009collapses}]
Let $\complex$ be a proper subcomplex of the $d$-pseudomanifold $\manifold$. If the dimension of $\complex$ is equal to $d$, then necessarily there exists a free $d$-pair for $\complex$. In other words, the dimension of an ultimate
$d$-collapse of $\complex$ is necessarily $d-1$.
\label{prop:thin}
\end{Proposition}
 
Following Prop.~\ref{prop:thin}, we say that an ultimate $d$-collapse of $\complex\subset\manifold$ is \emph{thin}.

Let $x \in \manifold$, the star of $x$ (in $\manifold$), denoted by $\STAR(x)$, is the set of all simplices of $\manifold$ that include $x$, i.e., $\STAR(x)=\{y \in \manifold \;|\; x \subseteq y\}$. If $A$ is a subset of $\manifold$, the set $\STAR(A) = \cup_{x \in A}\STAR(x)$ is called the \emph{star} of $A$ (in $\manifold$). A set $A$ of simplices of $\manifold$ is a star (in $\manifold$) if $A = star(A)$.

\subsection{Simplicial stacks}

 Let $\myfunction$ be a mapping $\manifold \rightarrow \Zeals$. For any face $\sigma$ of $\manifold$, the value $\myfunction(\sigma)$ is called the \emph{altitude} of $\myfunction$ at $\sigma$. For $k \in \Zeals$, the \emph{$k$-section} of $\myfunction$, denoted by $[\myfunction \geq k]$ is equal to $\{\sigma \in \manifold \; | \; \myfunction(\sigma) \geq k\}$.  A \emph{simplicial stack} $\myfunction$ on $\manifold$ is a map from $\manifold$ to $\Zeals$ which satisfies that any of its $k$-section is a (possibly empty) simplicial complex. In other words, a map $\myfunction$ is a simplicial stack if, for any two faces $\sigma$ and $\tau$ of $\manifold$ such that $\sigma \subseteq \tau$, $\myfunction(\sigma) \geq \myfunction(\tau)$.

We say that a subset $A$ of $\manifold$ is a \emph{minimum} of $\myfunction$ at altitude $k \in \Zeals$ when $A$ is a connected component of $[\myfunction \leq k] := \{ \sigma \in \manifold \; | \; \myfunction(\sigma) \leq k\}$ and $A \cap [F \leq k - 1] = \emptyset$. In the following, we denote by $\minima(\myfunction)$ the union of all minima of $\myfunction$. We note that, if $\myfunction$ is a simplicial stack, then $\minima(\myfunction)$ is a star.
The \emph{divide} of a simplicial stack $\myfunction$ is the set of all faces of $\manifold$ which do not belong to any minimum of $\myfunction$. Note that since $\minima(\myfunction)$, is a star, the divide is a simplicial complex.

Let $\sigma$ be any face of $\manifold$. When $\sigma$ is a free face for $[\myfunction \geq \myfunction(\sigma)]$, we say that $\sigma$ is a \emph{free face for $\myfunction$}. If $\sigma$ is a free face for $\myfunction$, there exists a unique face $\tau$ in $[\myfunction \geq \myfunction(\sigma)]$ such that $(\sigma,\tau)$ is a free pair for $[\myfunction \geq \myfunction(\sigma)]$, and we say that $(\sigma,\tau)$ is a \emph{free pair} for $\myfunction$. Let $(\sigma,\tau)$ be a free pair for $\myfunction$, then it is also a free pair for $[\myfunction \geq \myfunction(\sigma)]$. Thus, $\tau$ is a face of $[\myfunction \geq \myfunction(\sigma)]$, and we have $\sigma \subseteq \tau$. Therefore, we have $\myfunction(\tau) \geq \myfunction(\sigma)$ and $\myfunction(\tau) \leq \myfunction(\sigma)$ (since $\myfunction$ is a stack), which imply that $\myfunction(\tau) = \myfunction(\sigma)$. Let $\manifoldtwo \subseteq \manifold$, the \emph{indicator function} of $\manifoldtwo$, denoted by $\indicatortwo : \manifold \rightarrow \{0,1\}$, is the mapping such that $\indicatortwo(\sigma)$ is equal to $1$ when $\sigma$ belongs to $\manifoldtwo$ and is equal to $0$ when $\sigma$ belongs to $\manifold \setminus \manifoldtwo$. The \emph{lowering} of $\myfunction$ at $\manifoldtwo$ is the map $\myfunction - \indicatortwo$ from $\manifold$ into $\Zeals$. Let $(\sigma^{(d-1)},\tau^{(d)}$) be a free pair for $\myfunction$. The map $\myfunction - \indicatorST$ is called an \emph{elementary $d$-collapse of $\myfunction$}. Thus, this elementary $d$-collapse is obtained by subtracting $1$ to the values of $\myfunction$ at $\sigma$ and $\tau$. Note that the obtained mapping is still a simplicial stack. If a simplicial stack $\myfunctiontwo$ is the result of a sequence of elementary $d$-collapses on $\myfunction$, then we say that $\myfunctiontwo$ is a \emph{$d$-collapse} of $\myfunction$. If, furthermore, there is no free pair $(\sigma^{(d-1)},\tau^{d})$ for $\myfunctiontwo$, then $\myfunctiontwo$ is an \emph{ultimate $d$-collapse} of $\myfunction$.

\subsection{Watersheds of simplicial stacks}

Let $A$ and $B$ be two nonempty stars in $\manifold$.
We say that $B$ is an extension of $A$ if $A \subseteq B$, and if each connected component of $B$ includes exactly one connected component of $A$. We also say that $B$ is an {\em extension} of $A$ if $A = B = \emptyset$. Let $X$ be a subcomplex of the pseudomanifold $\manifold$ and let $Y$ be a collapse of $X$, then the complement of $Y$ in $\manifold$ is an extension of the complement of $X$ in $\manifold$. Let $A$ be a nonempty open set in a pseudomanifold $\manifold$ and let $X$ be a subcomplex of $\manifold$. We say that $X$ is a \emph{cut} for $A$ if the complement of $X$ is an extension of $A$ and if $X$ is minimal for this property. Observe that there can be several distinct cuts for a same open set $A$ and, in this case, these distinct cuts do not necessarily contain the same number of faces. 

Let $\pi = \langle x_0,\dots,x_{\ell} \rangle$ be a $d$-path in $\manifold$, and let $\myfunction$ be a function on $\manifold$.  We say that the $d$-path $\pi$ is \emph{descending} for $\myfunction$ if for any $i \in \{1,\dots,\ell\}$, $\myfunction(x_i) \leq \myfunction(x_{i-1})$.

Let $X$ be a subcomplex of the pseudomanifold $\manifold$. We assume that $X$ is a cut for $\minima(\myfunction)$. We say that $X$ is a \emph{watershed-cut} of $\myfunction$ if for any $x \in X$, there exists two descending paths $\pi_1 = \langle x, x_0, \dots,x_{\ell} \rangle$ and $\pi_2 = \langle x, y_0, \dots, y_m \rangle$ such that (1) $x_{\ell}$ and $y_m$ are simplices of two distinct minima of $\myfunction$; and (2) $x_i \not \in X$, $y_j \not \in X$, for any $i \in \{0,\dots,\ell\}$ and $j \in \{0,\dots,m\}$.

Several equivalent definitions of the watershed for pseudo-manifolds are given in~\cite{cousty2014collapses,cousty2009collapses}. 
Also, it was shown that a watershed-cut of $\myfunction$ is necessarily included in an ultimate $d$-collapses of $\myfunction$.
 Thus, by Prop.~\ref{prop:thin}, a watershed-cut is a thin divide.
 
 In this paper, we focus on a definition relying on combinatorial optimization, more precisely on the minimum spanning tree. For that, we need  a notion of “dual graph” of a pseudomanifold.

Starting from a $d$-pseudomanifold $\manifold$ valued by $\myfunction : \manifold \rightarrow \Zeals$, we define the {\em dual (edge-weighted) graph} of $\myfunction$ as the $3$-tuple $\graphmanifold=(V,E,\myfunctiongraph)$ whose vertex set $V$ is composed of the $d$-simplices of $\manifold$, whose edge set $E$ is composed of the pairs $\{\sigma,\tau\}$ such that $\sigma, \tau$ are $d$-faces of $\manifold$ and $\sigma \cap \tau$ is a $(d-1)$-face of $\manifold$, and whose edge weighting $\myfunctiongraph$ is made as follows: for two distinct $d$-faces $\sigma,\tau$ in $\manifold$ sharing a (d-1)-face of $\manifold$, $\myfunctiongraph(\{\sigma,\tau\}) = \myfunction(\sigma \cap \tau)$.

Let $A$ and $B$ be two non-empty subgraphs of the dual graph $\graphmanifold$ of $\myfunction$. We say that $B$ is a forest relative to $A$ when 
\begin{itemize}
    \item[(1)] $B$ is an extension of $A$; and
    \item[(2)] for any extension $C \subseteq B$ of $A$, we have $C = B$ whenever $B$ and $C$ share the same vertices.
\end{itemize} 
Informally speaking, the second condition imposes that we cannot remove any edge from $B$ while keeping an extension of $A$ that has the same vertex set as $B$. We say that $B$ is a \emph{spanning forest relative to $A$} for $\graphmanifold$ if $B$ is a forest relative to $A$ and if $B$ and $\graphmanifold$ share the same vertices.

The \emph{weight} of $A$ is defined as: $\myfunctiongraph(A) := \sum_{u \in E(A)} \myfunctiongraph(u)$. We say that $B$ is a \emph{minimum spanning forest (MSF)} relative to $A$ for $\myfunctiongraph$ if $B$ is a spanning forest relative to $A$ for $\myfunctiongraph$ and if the weight of $B$ is less than or equal to the weight of any other spanning forest relative to $A$ for $\myfunctiongraph$.

Let $A$ be a subgraph of $\graphmanifold$, and let $X$ be a set of edges of $\graphmanifold$. We say that $X$ is an \emph{MSF cut} for $A$ if there exists an MSF $B$ relative to $A$ such that $X$ is the set of all edges of $\graphmanifold$ adjacent to two distinct connected components of $B$.

In the following, if $S$ is a set of $(d-1)$-faces of $\manifold$, we set $\Edges(S) =\left\{  \{\sigma,\tau\} \in E(\graphmanifold) \; | \; \sigma \cap \tau \in S \right\}$. The {\em dual graph of the minima} of $\myfunction$ is the graph whose vertex set is the set $M$ of $d$-faces of the minima of $\myfunction$ and whose edge set is composed of the edges of $\graphmanifold$ linking two elements of $M$.

\begin{Th}[Theorem 16 p. 10~\cite{cousty2009collapses}]
\label{th.computation.WS.using.dual.graph}
Let $X$ be a set of $(d-1)$-faces of $\manifold$, and let $\myfunction:\manifold\rightarrow\Zeals^+$ be a simplicial stack. The complex resulting from the closure of $X$ is a watershed-cut of $\myfunction$ if, and only if, $\Edges(X)$ is a MSF cut for the dual graph of the minima of $\myfunction$. 
\end{Th}

In other words, to compute the watershed of a stack $\myfunction$, it is sufficient to compute in $\graphmanifold$ a MSF cut relative to the graph associated with the minima of $\myfunction$. Different algorithms for computing MSF cuts are detailed in~\cite{cousty2009watershed,cousty2010watershed}.

\subsection{Basic Discrete Morse functions}
We rely here on the formalism presented in~\cite{scoville2019discrete}, with results from Forman and Benedetti.
A function $\myfunction : A \rightarrow B$ is said to be $2-1$ when, for every $b \in B$, there exist at most two values $a_1,a_2 \in A$ such that $\myfunction(a_1) = \myfunction(a_2) = b$. Let $\complex$ be a simplicial complex. A function $\myfunction : \complex \rightarrow \Zeals$ is called \emph{weakly increasing} if $\myfunction(\sigma) \leq \myfunction(\tau)$ whenever the two faces $\sigma,\tau$ of $\complex$ satisfy $\sigma \subseteq \tau$. 

A \emph{basic discrete Morse function} $\myfunction : \complex \rightarrow \Zeals$ is a weakly increasing function which is $2-1$ and satisfies the property that if $\myfunction(\sigma) = \myfunction(\tau)$, then $\sigma \subseteq \tau$ or $\tau \subseteq \sigma$. 

Let $\myfunction : \complex \rightarrow \Zeals$ be a basic discrete Morse function.
A simplex $\sigma$ of $\complex$ is said to be \emph{critical} when $\myfunction$ is injective on $\sigma$. Otherwise, $\sigma$ is called \emph{regular}. When $\sigma$ is a critical simplex, $\myfunction(\sigma)$ is called a \emph{critical value}. If $\sigma$ is a regular simplex, $\myfunction(\sigma)$ is called a \emph{regular value}. 

Discrete Morse functions are more general than basic discrete Morse functions. A discrete Morse function (DMF) $\myfunction$ on $\complex$ is a function from $\myfunction:\complex\rightarrow\mathbb{Z}$ such that for every $p$-simplex $\sigma\in\complex$, we have
\begin{equation}
|\{ \tau^{(p-1)} \subset \sigma \;|\; \myfunction(\tau) \geq \myfunction(\sigma) \}| \leq 1
\end{equation}
and
\begin{equation}
|\{ \tau^{(p-1)} \supset \sigma \;|\; \myfunction(\tau) \leq \myfunction(\sigma) \}| \leq 1.
\end{equation}
However, to each discrete Morse function, there exists a basic discrete Morse function which is equivalent in the following sense (see Th.~\ref{th.classes.morse} and Prop.~\ref{th.basic.equivalence}). Two discrete Morse functions $\myfunction,\myfunction'$ defined on the same simplicial complex $\complex$ are said to be \emph{Forman-equivalent} when for any two faces $\sigma^{(p)},\tau^{(p+1)} \in \complex$ satisfying $\sigma \subset \tau$, $\myfunction(\sigma) < \myfunction(\tau)$ if and only if $\myfunction'(\sigma) < \myfunction'(\tau)$. Hence, in this paper, we focus on basic discrete Morse functions.

Let $\myfunction$ be a basic discrete Morse function on $\complex$. The \emph{(induced) gradient vector field} (GVF) $\gradient$ of $\myfunction$ is defined by 
\begin{equation}
\gradient(\myfunction) := \left\{ (\sigma^{(p)},\tau^{(p+1)}) \; | \; \sigma, \tau \in \complex \; , \; \sigma \subset \tau \; , \; \myfunction(\sigma) \geq \myfunction(\tau)\right\}.
\end{equation}
If $(\sigma,\tau)$ belongs to $\gradient(\myfunction)$, then it is called a \emph{vector} (for $\myfunction$) whose $\sigma$ is the \emph{tail} and $\tau$ is the \emph{head}. The vector $(\sigma,\tau)$ is sometimes denoted by $\overrightarrow{\sigma \tau}$.

Let $\complex$ be a simplicial complex. A \emph{discrete vector field} $V$ on $\complex$ is defined by
\begin{equation}
    V := \{ (\sigma^{(p)},\tau^{(p+1)}) \; | \; \sigma\subset \tau, \text{each simplex of $\complex$ is in at most one pair}\}
\end{equation}
Naturally, every GVF is a discrete vector field. 

Let $V$ be a discrete vector field on a simplicial complex $\complex$. 
A \emph{gradient path} is a sequence of simplices: $(\tau_{-1}^{(p+1)},) \sigma_{0}^{(p)},\tau_{0}^{(p+1)}, \sigma_{1}^{(p)},\tau_{1}^{(p+1)}, \dots,  \sigma_{k-1}^{(p)},\tau_{k-1}^{(p+1)}, \sigma_{k}^{(p)},$ of $\complex$, beginning at either a critical simplex $\tau_{-1}^{(p+1)}$ or a regular simplex $\sigma_{0}^{(p)}$, such that $(\sigma_{\ell}^{(p)},\tau_{\ell}^{(p+1)})$ belongs to $V$ and $\tau_{\ell-1}^{(p+1)} \supset \sigma_{\ell}^{(p)}$ for $0 \leq \ell \leq k - 1$. If $k \neq 0$, then this path is said to be \emph{non-trivial}. Note that the last simplex does not need to be in a pair in $V$. A gradient path is said to be \emph{closed} if $\sigma_{k}^{(p)} = \sigma_{0}^{(p)}$.

\begin{Th}[Theorem 2.51 p.61 of~\cite{scoville2019discrete}]
\label{th.no.cycle.in.gradient.vector.field}
A discrete vector field is the GVF of a discrete Morse function if, and only if, this discrete vector field contains no non-trivial closed gradient paths.
\end{Th}

\begin{Th}[Theorem 2.53 p.62 of~\cite{scoville2019discrete}]
\label{th.classes.morse}
Two discrete Morse functions defined on a same complex $\complex$ are Forman-equivalent if, and only if, they induce the same GVF. A consequence is that any two Forman-equivalent discrete Morse functions defined on a simplicial complex have the same critical simplices.
\end{Th}

\begin{Proposition}
\label{th.basic.equivalence}
If $\myfunction$ is a discrete Morse function, there exists $\myfunction'$ a basic discrete Morse function that is Forman-equivalent to $\myfunction$. 
\end{Proposition}

\newcommand{\KGRAD}{$k$-\mathrm{Grad}\xspace}

\elimine{
\begin{algorithm}[tbp]
\caption{Building a basic DMF Forman-equivalent to a given DMF.}
\label{algo.makingbasic}
\dontprintsemicolon
\Begin{
\lcomment{$\myfunction$ is a  DMF}\;
\lcomment{$\gradient$ is the GVF of $\myfunction$}\;
$cpt \setto 0$\;
\For{$k$ \setto $0$ to $d$}{
\lcomment{Valuating critical faces}\;
$C \setto \text{critical faces of $\myfunction$ of dimension $k$}$\;
\While{$C \neq \emptyset$}{
$h \setto Pop(C)$\;
$\myfunction'(h) = cpt$\;
$cpt++$\;
}
\If{$k < d$}{
\lcomment{Valuating paired faces}\;
$\KGRAD \setto \{ (h,h') \; ; (h,h') \in \gradient, \dim(h) = k, \dim(h') = k+1 \}$\;
\For{$S \in \connectedcomponent(\KGRAD)$}{
\For{$depth \setto Depth(S) \mbox{~to~} 0$}{
$ S_p \setto \{(h,h') \in S ; Depth((h,h'),S) == depth\}$\;
\While{$S_p \neq \emptyset$}{
$(h,h') \setto Pop(S_p)$\;
$\myfunction'(h) = cpt$\;
$\myfunction'(h') = cpt$\;
$cpt++$\;
}
}
}
}
}
\return $\myfunction'$ \tcc{A basic DMF}
}
\end{algorithm}
}

Proposition \ref{th.basic.equivalence} is a consequence of \cite[Proposition 4.16]{scoville2019discrete}.
Starting from a DMF and computing its GVF, it is possible (by correctly ordering all the simplices) to compute a basic DMF Forman-equivalent to it; such an algorithm preserves the GVF. A precise algorithm, together with a proof of Prop~\ref{th.basic.equivalence} relying on this algorithm, will be provided in an extended version of this paper.

\section{A class of simplicial stack equivalent to Morse functions}
\label{sec.equiv}
Simplicial stacks are weakly decreasing. We call {\em basic simplicial stack}, a simplicial stack $\myfunction$ that is $2-1$ and satisfies the property that if $\myfunction(\sigma) = \myfunction(\tau)$, then $\sigma \subseteq \tau$ or $\tau \subseteq \sigma$.
The proof of the following is straightforward.
\begin{Proposition}
Let $\myfunction$ be a function defined on $\manifold$. Then $\myfunction$ is a basic simplicial stack if and only if $- \myfunction$ is a basic discrete Morse function.
\label{propo.StackToMorseByMinus}
\end{Proposition}

Hence, all properties of basic discrete Morse functions hold true for basic simplicial stacks, and conversely. In the sequel of this paper, we exemplify that fact with gradient vector fields.

\elimine{
\begin{figure}[tbp]
    \centering
    \includegraphics[width=0.6\linewidth]{Figures/gradientIsTheCode.png}
    \caption{The GVF encodes the possible collapses in both basic discrete Morse functions and basic simplicial stacks.}
        \label{fig:gradientIsTheCode}
\end{figure}
}

Relying on Prop.~\ref{propo.StackToMorseByMinus}, we define the {\em gradient vector field of a basic simplicial stack} $\myfunction$ as the GVF of the DMF $-\myfunction$ it corresponds to.
 
As stated in Th.~\ref{th.classes.morse}, two basic DMF's are Forman-equivalent if, and only if, they induce the same GVF. In other words, at each GVF corresponds a class of DMF's. Using  Proposition~\ref{propo.StackToMorseByMinus}, we have a bijection between the space of basic DMF's and the space of basic simplicial stacks. This leads to the following corollary:

\begin{Coro}
If $\myfunction$ is a basic DMF defined on $\manifold$, there exists a class $\mathbb{BD}$ of basic DMF's and a class $\mathbb{SS}$ of basic simplicial stacks, bijective to $\mathbb{BD}$, such that each $\myfunctiontwo$ in one of those classes has the same gradient vector field as the one of $\myfunction$. 
\end{Coro}

\section{The minimum spanning forest of a stack and the GVF}
\label{sec.theproof}

\subsection{The forest induced by a GVF}

\newcommand{\SETTWOCOFACES}{\mathfrak{CO}_2\xspace}

Let $\myfunction$ a basic simplicial stack. 
As any $k$-section of $\myfunction$ is a simplicial complex, and as $\myfunction$ is 2-1, we have the following proposition:

\begin{Proposition}
Let $\myfunction$ be a basic simplicial stack. We have:
\begin{enumerate}
    \item $\minima(\myfunction)$ is a set of simplices of dimension $d$.
    \item Each minimum of $\myfunction$ is made of a single simplex of $\minima(\myfunction)$.
    \item The set of edges of the dual graph of the minima is empty.
\end{enumerate}   
\label{prop:minima}
\end{Proposition}

\newcommand{\barycenter}{\mathit{barycenter}\xspace}
\elimine{
\begin{algorithm}[tb]
\caption{Computing the MSF from the gradient vector field.}
\label{algo.2}
\dontprintsemicolon
\Begin{
\lcomment{Computation of the minima.}\;
$V = (\manifold)_d$\;
$E = \{\}$\;
\For{$m_1,m_2 \in (\minima(\myfunction))_d$}{
    \If{$m_1 \cap m_2 \in (\manifold)_{d-1}$}{
        $\push(m_1 \cap m_2, E)$\;
        }
    }
\lcomment{Computation of the edges of the MSF induced by the GVF.}\;
\For{$\overrightarrow{ab} \in \gradient$}{
    $c \setto \opp_{a}(b)$\;
    $\push(cb, E)$\;
    }
}
\return (V,E) \tcc{The MSF}\;
\end{algorithm}
}

 Let $\gradient$ be the GVF of $\myfunction$. Let $\VECTEURAB$ be a vector of $\gradient$ such that $dim(a) = d-1$.
 Since $\manifold$ is a pseudomanifold, the face $a$ is included in two $d$-faces, the face $b$ and another $d$-face $c$.
 We write $[\VECTEURAB] = \{ \{b\}, \{c\}, \{b,c\}\}$ and we consider the graph:
\begin{equation}
    {\cal G}(\gradient) = \cup \{[\VECTEURAB] \;|\; \VECTEURAB \in \gradient, 
    dim(a) = d-1 \}.
\end{equation} 

Let $\EXT(\gradient)$  be the union of ${\cal G}(\gradient)$ and of ${\cal G}(\minima(\myfunction))$, where ${\cal G}(\minima(\myfunction))$ is the dual graph of the minima of $\myfunction$.

\begin{Proposition}
The graph $\EXT(\gradient)$ is a spanning forest relative to the dual graph of the minima of $\myfunction$. 
\label{ExtGradIsSpanning}
\end{Proposition}

The proof of this proposition relies on the following fact: any critical simplices of $\myfunction$ that is not a minimum of $\myfunction$ is of dimension strictly lower than $d$. 

\elimine{
To prove this proposition, we need the following lemma.
\begin{Lem}[Exclusion lemma, Lemma 2.24 of \cite{scoville2019discrete}]
Let $myfunction:\complex\rightarrow\Zeals$ be a discrete Morse function and $\sigma\in\complex$ be a regular $p$-simplex. Then exactly one of the two following conditions holds true:
\begin{itemize}
    \item[(i)] There exists $\tau^{(p+1)}\supset\sigma$ such that $\myfunction(\tau)\leq\myfunction(\sigma)$
    \item[(ii)] There exists $\nu^{(p-1)}\subset\sigma$ such that $\myfunction(\nu)\geq\myfunction(\sigma)$\end{itemize}
\end{Lem}
}

\smallskip
\myproof %
 We first show that $\EXT(\gradient)$ spans all vertices of the dual graph $\graphmanifold$: as $\EXT(\gradient)$ contains the dual graph of the minima, we only need to show that for any  $d$-face $\sigma$ of $\manifold$, $\sigma\not\in\minima(\myfunction)$, there is a pair $(\tau^{(d-1)},\sigma)$ of simplices in $\gradient$. 

\cite[Remark 2.42]{scoville2019discrete} states that, for any simplex $\tau$, exactly one of the following holds true:
\begin{itemize}
    \item [(i)] $\tau$ is the tail of exactly one vector
    \item [(ii)] $\tau$ is the head of exactly one vector
    \item [(iii)] $\tau$ is neither the tail nor the head of a vector; that is $\tau$ is critical
\end{itemize}

By Prop.~\ref{prop:minima}, item 2, each minimum of $\myfunction$ is made of a single simplex of $\minima(\myfunction)$.
By remark  \cite[Remark 2.42]{scoville2019discrete} above, it remains to show that, if $\sigma$ is not a minimum of $\myfunction$, it is regular, and hence the head of exactly one vector. As $\myfunction$ is a simplicial stack, its $k$-section for $k=\myfunction(\sigma)$ contains all the simplices $\nu$ such that $\nu\subset\sigma$. We have $\myfunction(\nu)\geq\myfunction(\sigma)$.  Because $\sigma$ is not  a minimum, there exists a simplex $\tau^{(d-1)}$ such that $\myfunction(\tau^{(d-1)})=\myfunction(\sigma)$ with $\tau^{(d-1)}\subset\sigma$. This implies that  $(\tau^{(d-1)},\sigma)\in\gradient$. 
 
 By Th.~\ref{th.no.cycle.in.gradient.vector.field}, ${\cal G}(\gradient)$ does not contain any closed $1$-path. Hence, $\EXT(\gradient)$ is a forest relative to the dual graph of the minima of $\myfunction$.
 \qed

Following Prop.~\ref{ExtGradIsSpanning}, we say in the sequel that $\EXT(\gradient)$ is the \emph{forest induced} by the GVF $\gradient$. 

\subsection{The forest induced by a GVF is the  MSF}

\begin{Proposition}
Let $\myfunction : \manifold \rightarrow \Zeals$ be a basic simplicial stack, and let $\gradient$ be the GVF of $\myfunction$. Then, any gradient-path $\pi = (\pi(k))_{k \in [0,N]}$ of the GVF is \emph{increasing}, that is, for any $k \in [0,N-1]$, $\myfunction(\pi(k)) \leq \myfunction(\pi(k+1))$.
\label{path.increasing.grad}
\end{Proposition}

\myproof Let $\pi$ some gradient-path  of $\gradient$, and let us assume without loss of generality, that $\pi(0)$ is a $d$-face of $\manifold$. We know that the $(d-1)$-face $\pi(2 k + 1)$ is paired with the $n$-face $\pi(2 k + 2)$ in $\gradient$ for any $k \in [0,(N-1)/2 - 1]$ ($N$ is odd), which means that $\myfunction(\pi(2 k + 1)) = \myfunction(\pi(2 k + 2))$. We also know that $\myfunction$ is a stack, and then $\myfunction$ decreases when we increase the dimension of the face, so for any $k \in [0,(N-1)/2 - 1]$, $\myfunction(\pi(2 k)) \leq \myfunction(\pi(2 k + 1))$. \qed

\begin{Lem}[MST Lemma~\cite{motwani1995randomized,cormen1999introduction}]
Let $\graph = (V,E,\myfunction)$ be some edge-weighted graph. Let $v \in V$ be any vertex in $\graph$. A minimum spanning tree for $\graph$ must contain an edge $vw$ that is a minimum weighted edge incident on $v$.
\label{propo.MSTlemma}
\end{Lem}

\begin{Th}
Let $\myfunction : \manifold \rightarrow \Zeals^{+}$ be a basic simplicial stack, and let $\gradient$ be the GVF of $\myfunction$.
The forest induced by $\gradient$ is the unique MSF relative to $\minima(\myfunction)$ of the dual graph of $\myfunction$.
\label{th.principal}
\end{Th}

\myproof 
By Prop.~\ref{ExtGradIsSpanning}, $\EXT(\gradient)$ is a spanning forest relative to $\minima(\myfunction)$, the minima of $\myfunction$. As $\myfunction$ is a basic simplicial stack, hence 2-1, all edges of the dual graph $\graphmanifold = (V,E,\myfunctiongraph)$ of $\myfunction$ have a unique weight, and the MSF of the dual graph $\graphmanifold$ is unique. It remains to prove that the induced forest is of minimum cost.

Since gradients do not exist on minima, let us consider a $d$-simplex $\sigma\in V\setminus\minima(\myfunction)$. Then, by Prop.~\ref{ExtGradIsSpanning}, there exists exactly one vector in $\gradient$, which can be written $(\tau \cap \sigma, \sigma)$, with $\tau\in V$. By the definition of $\gradient$, we have $\myfunction(\tau \cap \sigma) = \myfunction(\sigma)$.

Let $\theta \in V \setminus \{\tau\}$ some $d$-simplex such that $\{\tau,\theta\}$ belongs to $E$. Since $\myfunction$ is a simplicial stack, either the $(d-1)$-face $\tau \cap \theta$ is critical (and $\myfunction(\tau \cap \theta) > \max(\myfunction(\tau),\myfunction(\theta))$), or it is regular and $\tau \cap \theta$ is paired with $\theta$ in $\gradient$ (and $\myfunction(\theta) = \myfunction(\tau \cap \theta) > \myfunction(\tau)$ by Proposition~\ref{path.increasing.grad}). Therefore, $\{\sigma,\tau\}$ is the lowest cost edge incident to $\tau$:
\begin{equation}
\myfunctiongraph(\{\sigma,\tau\}) = \myfunction(\tau \cap \sigma) = \myfunction(\sigma) < \min\{\myfunctiongraph(\{\theta,\tau\}) \; ; \; \{\theta,\tau\} \in E,\ \theta \neq \sigma\}
\end{equation}
and thus belongs to the MST of $\myfunction$ by Lemma~\ref{propo.MSTlemma}.

As by Prop~\ref{ExtGradIsSpanning}, the induced forest is a spanning forest relative to the dual graph of the minima of $\myfunction$, it is then the minimum spanning tree of the dual graph relative to the minima of $\myfunction$, which concludes the proof.
\qed

\begin{figure}[tb]
    \centering
        \includegraphics[width=.75\linewidth]{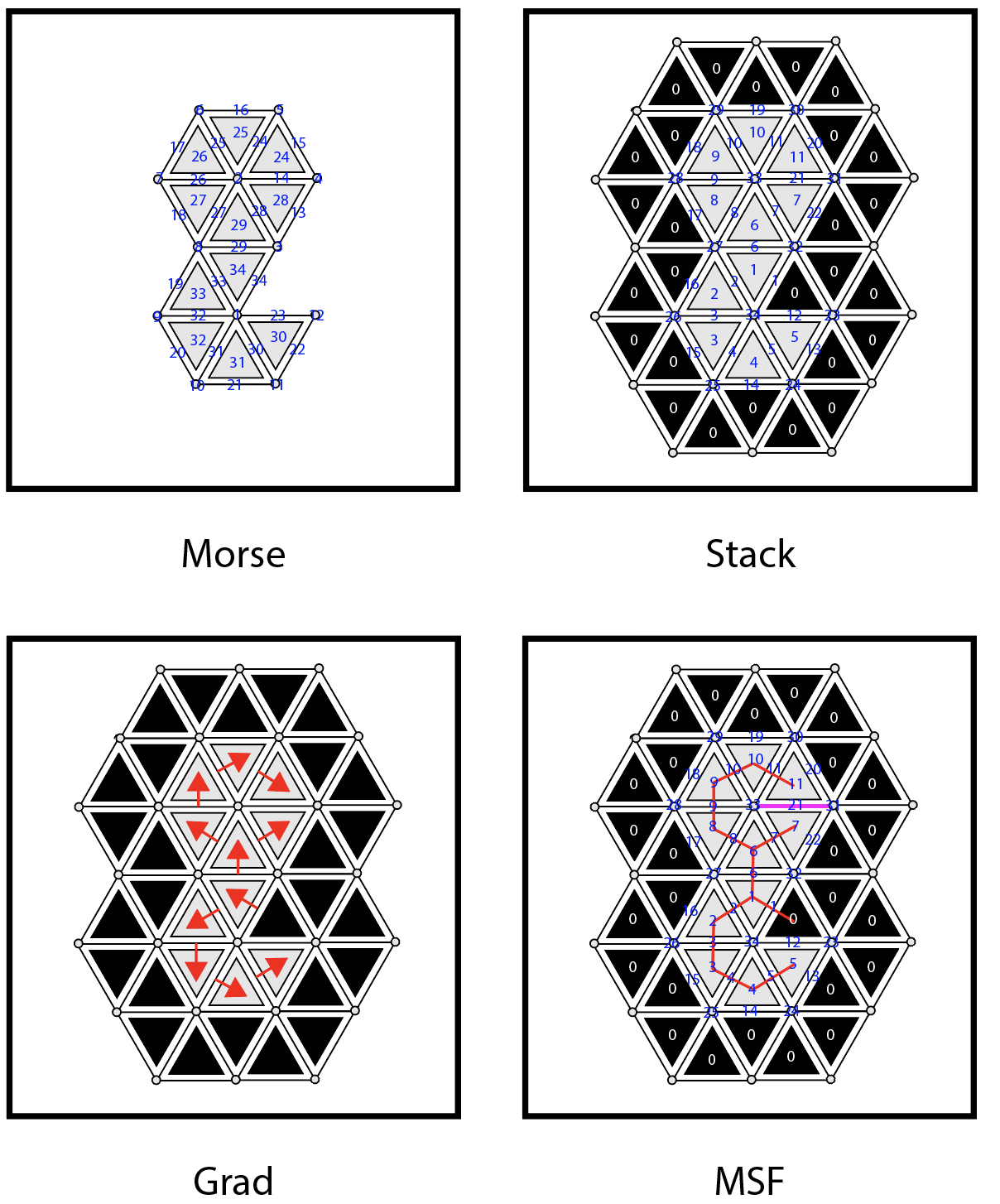}
    \caption{Starting from a Morse function, we obtain its equivalent simplicial stack up to the minus sign. For simplicity, the simplicial stack is valued by 0 on all $d$-simplices at the border. Then, we deduce the GVF of the initial Morse function and its MSF.  This illustrates that \textbf{the MSF is the forest induced by the  GVF} of both a discrete Morse function and the corresponding simplicial stack.}
    \label{fig:computingMSFofMorse}
\end{figure}

A summary of this result is depicted in Figure~\ref{fig:computingMSFofMorse}, which shows a piece of a pseudomanifold of dimension $2$.

\begin{figure}[tb]
    \centering
    \includegraphics[width=0.9\linewidth]{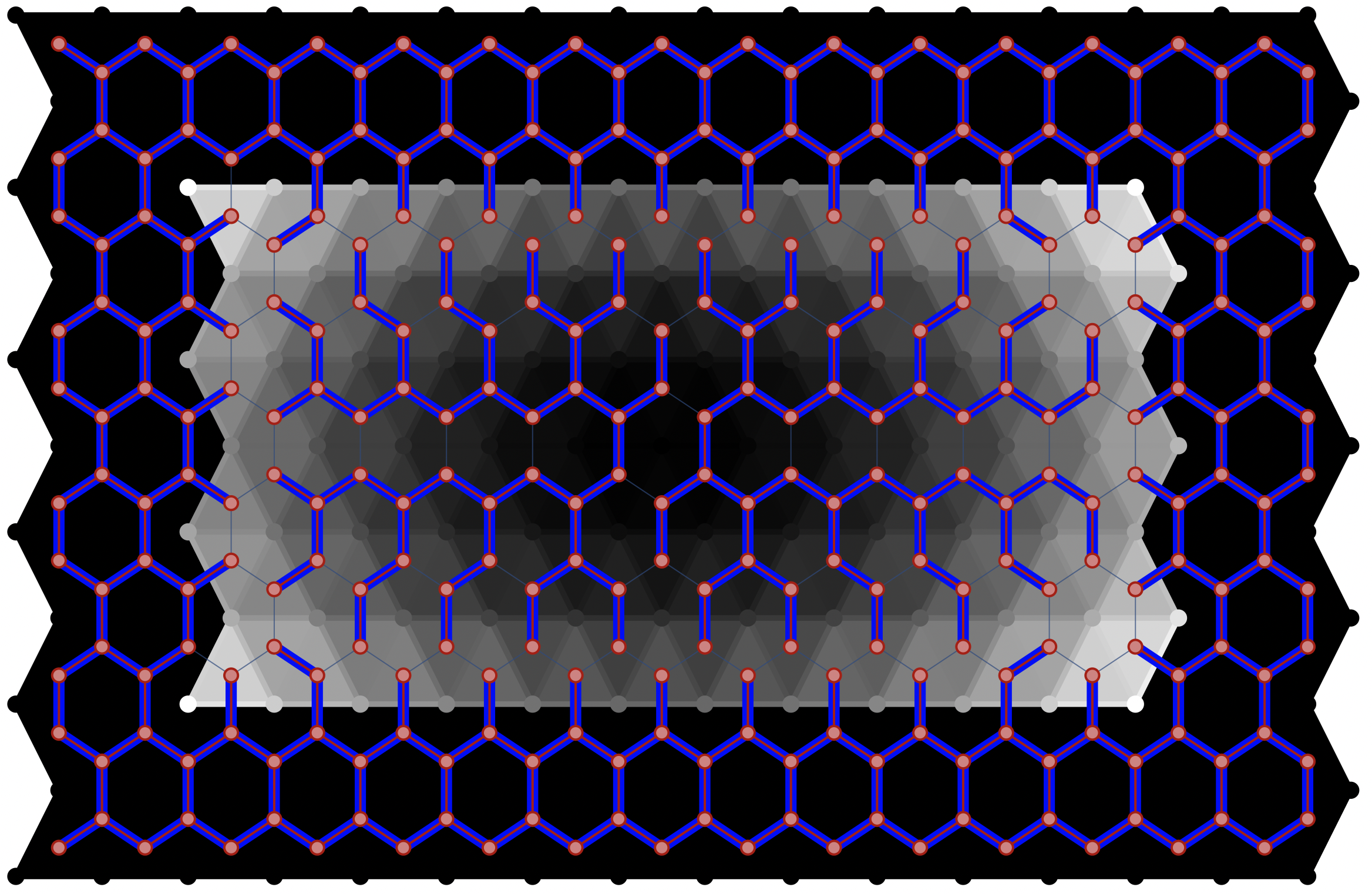}
    \caption{Illustration of the watershed-cut obtained from the GVF of a discrete Morse function: we have a partition of the pseudo-manifold, such that each tree in the forest is a basin of the watershed-cut. We also highlight the equality between the forest induced by the GVF (in blue) and the MSF  of the dual graph (in red).
    }
    \label{fig:comparison}
\end{figure}

Using Th.~\ref{th.computation.WS.using.dual.graph} and Th.~\ref{th.principal}, we can conclude that the cut of the forest induced by the GVF is also a watershed-cut. This leads to the following corollary.

\begin{Coro}
Let $\myfunction : \manifold \rightarrow \Zeals^+$ be a basic simplicial stack. Then, the watershed-cut of $\myfunction$ is provided equivalently by the MSF of $\myfunction$ or by the GVF of $\myfunction$.
\end{Coro}

Fig.~\ref{fig:comparison} illustrates this corollary: each tree of the induced forest is a connected component of the dual graph, called a catchment basin of the watershed-cut.

\section{Conclusion}
\label{sec.conclusion}

In this paper, we highlight some links between several notions that exist in Discrete Topology and in Mathematical Morphology:
\begin{itemize}
    \item discrete Morse functions are equivalent, under some constraints, to simplicial stacks;
    \item gradient vector fields in the Morse sense are applicable to simplicial stacks;
    \item and the gradient vector field of a simplicial stack induces the Minimum Spanning Forest of its dual graph, leading to watershed-cuts. 
\end{itemize} 
In the extended version of this paper,  we will relax the constraints for the equivalence between discrete Morse function and simplicial stacks, and we will show how to use the watershed to define a purely discrete version of the well-known Morse-Smale complex \cite{edelsbrunner2003morse}.

In the future, we will continue looking for strong relations linking Discrete Morse Theory and Mathematical Morphology, with the goal of using morphological tools for topological data analysis. We also aim at making clearer the relation between discrete topology and discrete Morse theory, following \cite{robins2011theory} that was inspired by \cite{couprie2008new}. 

\subsubsection*{Acknowledgements}
The authors would like to thank both Julien Tierny and Thierry Géraud, for many insightful discussions.

\bibliographystyle{splncs04}
\bibliography{article}

\end{document}